\documentstyle[12pt,epsf]{article}
\begin{document}
\topmargin -2truecm
\oddsidemargin 0truecm
\evensidemargin 0truecm
\textwidth 16truecm
\textheight 23truecm
\renewcommand{\thefootnote}{\arabic{footnote}}
%\newcommand{\be}{\begin{equation}}
%\newcommand{\ee}{\end{equation}}
%\input{tcilatex}
%\begin{document}
\def\non{\nonumber}
\def\be{\begin{eqnarray}}
\def\en{\end{eqnarray}}
\def\hep{\hat{\varepsilon}}
\def\la{\langle}
\def\ra{\rangle}
\def\pr{Phys. Rev.~}
\def\prl{Phys. Rev. Lett.~}
\def\pl{Phys. Lett.~}
\def\np{Nucl. Phys.~}
\def\zp{Z. Phys.~}

\thispagestyle{empty}
\begin{center}
{\Large {\bf Charge radii of light and heavy mesons}}\\
\vspace{1.7cm}
{\bf Chien-Wen Hwang \\[1cm]
{\em Department of Physics, National Tsing Hua University,
Hsinchu 300, Taiwan} \\[0.5cm]
{\small{\em e-mail: v00570@phys.nthu.edu.tw}}} \\[0pt]
\end{center}

\vspace{2cm}

\bigskip

%%%%%%%%%%%%%%%%%%%%%%%%%%%%%%abstract%%%%%%%%%%%%%%%%%%%%%%%%%%%%%%%
\begin{abstract}
We calculate the electromagnetic (EM) form factors of the pseudoscalar mesons in the light-front framework. Specifically, these form factors are extracted from the relevant matrix elements directly, instead of choosing the Breit frame. The results show that the charge radius of the meson is related to both the first and second longitudinal momentum square derivatives of the momentum distribution function. The static properties of the EM form factors and the heavy quark symmetry of the charge radii are checked analytically in the heavy quark limit. In addition, we use the Gaussia-type wavefunction to obtain the numerical results.
\end{abstract}
\thispagestyle{empty}
\newpage
%%%%%%%%%%%%%%%%%%%%%%%%%section one %%%%%%%%%%%%%%%%%%%%%%%%%%%%%%%%
\section{Introduction}
The understanding of the electromagnetic (EM) properties of hadrons is an important topic, and the EM form factors which are calculated using non-perturbative methods are the useful tool for this purpose. There have been numerous experimental [1-7] and theoretical studies [8-13] of the EM form factors of the light pseudoscalar mesons ($\pi$ and $K$). However, due to difficulties in the experiments, the EM form factors of light vector mesons ($\rho$ and $K^*$) have fewer investigations than their pseudoscalar counterparts \cite {HP,Simula}, even though they could provide much information about the bound-state dynamics. As for the EM form factors of heavy mesons (which containing one heavy quark), there are much fewer studies than the light ones. In the heavy hadron investigation, however, the heavy quark symmetry (HQS) \cite{HQS} is a fundamental and model-independent property. In this work, we will study the EM form factors of the light and heavy pseudoscalar mesons in the light-front framework. We will also check whether HQS is satisified or not among these EM properties of the heavy mesons.

The light front quark model (LFQM) is the only relativistic quark model in which a consistent and fully relativistic treatment of quark spins and the center-of-mass motion can be carried out. Thus it has been applied in the past to calculate various form factors [16-22]. This model has many advantages. For example, the light-front wavefunction is manifestly boost invariant as it is expressed in terms of the momentum fraction variables (in ``+" component) in analog to the parton distributions in the infinite momentum frame. Moreover, hadron spin can also be relativistically constructed by using the so-called Melosh rotation \cite{Melosh}. The kinematic subgroup of the light-front formalism has the maximum number of interaction-free generators including the boost operator which describes the center-of-mass motion of the bound state (for a review of the light-front dynamics and light-front QCD, see \cite{Zhang}). 

The paper is organized as follows. In Sec. 2, the basic theoretical formalism is given and the decay constant and the EM form factors are derived for pseudoscalar mesons. In Sec. 3,  we take the heavy quark limit to check whether HQS is satisfied or not. In Sec. 4, the numerical result are obtained by choosing the Gaussian-type wavefunction. Finally, a conclusion is given in Sec. 5.

%%%%%%%%%%%%%%%%%%%%%%%%%section two %%%%%%%%%%%%%%%%%%%%%%%%%%%%%%%%
\section{Framework}
A meson bound state consisting of a quark $q_1$ and
an antiquark $\bar q_2$ with a total momentum $P$
and spin $S$ can be written as
\begin{eqnarray}
        |M(P, S, S_z)\rangle
                &=&\int \{d^3p_1\}\{d^3p_2\} ~2(2\pi)^3 \delta^3(\tilde
                P-\tilde p_1-\tilde p_2)~\nonumber\\
        &&\times \sum_{\lambda_1,\lambda_2}
                \Psi^{SS_z}(\tilde p_1,\tilde p_2,\lambda_1,\lambda_2)~
                |q_1(p_1,\lambda_1) \bar q_2(p_2,\lambda_2)\rangle,
\end{eqnarray}
where $p_1$ and $p_2$ are the on-mass-shell light-front momenta,
\begin{equation}
        \tilde p=(p^+, p_\bot)~, \quad p_\bot = (p^1, p^2)~,
                \quad p^- = {m^2+p_\bot^2\over p^+},
\end{equation}
and
\begin{eqnarray}
        &&\{d^3p\} \equiv {dp^+d^2p_\bot\over 2(2\pi)^3}, \nonumber \\
        &&|q(p_1,\lambda_1)\bar q(p_2,\lambda_2)\rangle
        = b^\dagger_{\lambda_1}(p_1)d^\dagger_{\lambda_2}(p_2)|0\rangle,\\
        &&\{b_{\lambda'}(p'),b_{\lambda}^\dagger(p)\} =
        \{d_{\lambda'}(p'),d_{\lambda}^\dagger(p)\} =
        2(2\pi)^3~\delta^3(\tilde p'-\tilde p)~\delta_{\lambda'\lambda}.
                \nonumber
\end{eqnarray}
In terms of the light-front relative momentum
variables $(x, k_\bot)$ defined by
\begin{eqnarray}
        && p^+_1=(1-x) P^+, \quad p^+_2=x P^+, \nonumber \\
        && p_{1\bot}=(1-x) P_\bot+k_\bot, \quad p_{2\bot}=x P_\bot-k_\bot,
\end{eqnarray}
the momentum-space wavefunction $\Psi^{SS_z}$
can be expressed as
\begin{equation}
        \Psi^{SS_z}(\tilde p_1,\tilde p_2,\lambda_1,\lambda_2)
                = R^{SS_z}_{\lambda_1\lambda_2}(x,k_\bot)~ \phi(x, k_\bot), \label{momentumspace}
\end{equation}
where $\phi(x,k_\bot)$ describes the momentum distribution of the
constituents in the bound state, and $R^{SS_z}_{\lambda_1\lambda_2}$
constructs a state of definite spin ($S,S_z$) out of light-front
helicity ($\lambda_1,\lambda_2$) eigenstates.  Explicitly,
\begin{equation}
        R^{SS_z}_{\lambda_1 \lambda_2}(x,k_\bot)
                =\sum_{s_1,s_2} \langle \lambda_1|
                {\cal R}_M^\dagger(1-x,k_\bot, m_1)|s_1\rangle
                \langle \lambda_2|{\cal R}_M^\dagger(x,-k_\bot, m_2)
                |s_2\rangle
                \langle {1\over2}s_1;
                {1\over2}s_2|S,S_z\rangle, \label{RR}
\end{equation}
where $|s_i\rangle$ are the usual Pauli spinors,
and ${\cal R}_M$ is the Melosh transformation operator \cite{Melosh}:
\begin{equation}
        {\cal R}_M (x,k_\bot,m_i) =
                {m_i+x M_0+i\vec \sigma\cdot\vec k_\bot \times \vec n
                \over \sqrt{(m_i+x M_0)^2 + k_\bot^2}}, \label{Melosh2}
\end{equation}
with $\vec n = (0,0,1)$, a unit vector in the $z$-direction, and
\begin{equation}
        M_0^2={ m_1^2+k_\bot^2\over (1-x)}+{ m_2^2+k_\bot^2\over x}.
\label{M0}
\end{equation}
In practice, it is more convenient to use the covariant form for
$R^{SS_z}_{\lambda_1\lambda_2}$ \cite{Jaus}:
\begin{equation}
        R^{SS_z}_{\lambda_1\lambda_2}(x,k_\bot)
                ={\sqrt{p_1^+p_2^+}\over \sqrt{2} ~{\widetilde M_0}}
        ~\bar u(p_1,\lambda_1)\Gamma v(p_2,\lambda_2), \label{covariant}
\end{equation}
where
\begin{eqnarray}
        &&{\widetilde M_0} \equiv \sqrt{M_0^2-(m_1-m_2)^2}, \nonumber\\
        &&\Gamma=\gamma_5 \qquad ({\rm pseudoscalar}, S=0). \non 
\end{eqnarray}
We normalize the meson state as
\begin{equation}
        \langle M(P',S',S'_z)|M(P,S,S_z)\rangle = 2(2\pi)^3 P^+
        \delta^3(\tilde P'- \tilde P)\delta_{S'S}\delta_{S'_zS_z}~,
\label{wavenor}
\end{equation}
so that the normalization condition of the momentum distribution function can be obtained
\begin{equation}
        \int \{dx\}~|\phi(x,k_\bot)|^2 = 1,
\label{momnor}
\end{equation}
where
\be
\{dx\}\equiv {dx\,d^2k_\bot\over 2(2\pi)^3}\non
\en
In principle, the momentum distribution amplitude
$\phi(x,k_\bot)$ can be obtained by solving the light-front
QCD bound state equation \cite{Zhang}.
However, before such first-principles
solutions are available, we would have to be contented with
phenomenological amplitudes.  One example that has been often
used in the literature for heavy mesons is the Gaussian-type wavefunction,
\begin{equation}
        \phi(x,k_\bot)_{\rm G}={\cal N} \sqrt{{dk_z\over dx}}
        ~{\rm exp}\left(-{\vec k^2\over 2\omega^2}\right),
        \label{gauss}
\end{equation}
where ${\cal N}=4(\pi/\omega^2)^{3/4}$ and $k_z$ is of the internal momentum
$\vec k=(\vec{k}_\bot, k_z)$, defined through
\begin{equation}
1-x = {e_1-k_z\over e_1 + e_2}, \qquad
x = {e_2+k_z \over e_1 + e_2},
\end{equation}
with $e_i = \sqrt{m_i^2 + \vec k^2}$. We then have 
\be
M_0=e_1 + e_2,~~~~k_z = \,{xM_0\over 2}-{m_2^2+k_\perp^2 \over 2 xM_0},
\label{kz}
\en
and
\begin{equation}
        {{dk_z\over dx}} = \,{e_1 e_2\over x(1-x)M_0},
\end{equation}
which is the Jacobian of transformation from $(x, k_\bot)$ to
$\vec k$.
%%%%%%%%%%%%%%%%%%%%%%%%%subsection A
\subsection{Decay Constants}
The decay constant of a pseudoscalar meson $P(q_1\bar{q}_2)$ is defined by
\be
\la 0|A_\mu|P(p)\ra=\,i\,f_P\,p_\mu, \label{dcd18}
\en
where $A_\mu$ is the axial-vector current. It can be evaluated using the light-front wavefunction given by (\ref{gauss})
\be
\la 0|\bar{q}_2\gamma^+\gamma_5q_1|P\ra &=& \int \{d^3p_1\}\{d^3p_2\}
2(2\pi)^3\delta^3(\tilde p-\tilde p_1-\tilde p_2)\phi_P(x,k_\perp)R^{00}_{\lambda_1\lambda_2}
(x,k_\perp)   \non \\
&& \times\,\la 0|\bar{q}_2\gamma^+\gamma_5q_1|q_1\bar{q}_2\ra.
\en
Since $\widetilde{M}_0\sqrt{x(1-x)}=\sqrt{{\cal A}^2+k^2_\perp}$,
it is straightforward to show that
\be
f_P=\,4{\sqrt{3}\over{\sqrt{2}}}\int \{dx\}\,{\phi_P(x,
k_\perp)\over{\sqrt{{\cal A}^2+k_\perp^2}}}\,{\cal A}, \label{fp}
\en
where ${\cal A}=m_1x+m_2(1-x)$. Note that the factor $\sqrt{3}$ in (\ref{fp}) arises from the color factor implicitly in the meson wavefunction.

%%%%%%%%%%%%%%%%%%%%%%%%%subsection B
\subsection{Electromagnetic Form Factors}
The EM form factor of a pseudoscalar meson $P$, $F_P(Q^2)$, is determined by the scattering of one virtual photon and one meson. It describes the deviation from the point-like structure of the mesonch, and is a function of the square of the photon momentum $Q$. Here we consider the momentum of the virtual photon in space-like region, so it is always possible to orient the axes in such a manner that $Q^+=0$. Thus the EM form factor is determined by the matrix element
\be
\la P(P')|J^+|P(P)\ra = e~F_P(Q^2) (P+P')^+, \label{FPdef}
\en
where $J^\mu = \bar q e_q e\gamma^\mu q$ is the vector current, $e_q$ is the charge of quark $q$ in $e$ unit, and $Q^2=-(P'-P)^2\geq 0$. With LFQM, $F_P$ can be extracted by Eq. (\ref{FPdef})
\be
F_P(Q^2)&=&e_{q_1}\int \{dx\} {\phi_P(x,k_\perp)\over{\sqrt{{\cal A}^2+k^2_\perp}}}{\phi_{P'}(x,k'_\perp)\over{\sqrt{{\cal A}^2+k'^2_\perp}}}\left[{\cal A}^2+k_\perp\cdot k'_\perp\right]\non \\
&+&e_{\bar {q_2}}\int \{dx\}{\phi_P(x,k_\perp)\over{\sqrt{{\cal A}^2+k^2_\perp}}}{\phi_{P'}(x,k''_\perp)\over{\sqrt{{\cal A}^2+k''^2_\perp}}}\left[{\cal A}^2+k_\perp\cdot k''_\perp\right], \label{FPgeneral}
\en
where $k'_\perp=k_\perp+x Q_\perp$, $k''_\perp=k_\perp-(1-x) Q_\perp$. From Eqs. (\ref{RR}), (\ref{Melosh2}), and (\ref{covariant}), it is understandable that the term $\sqrt{{\cal A}^2+k_\perp^2}$ comes from the Melosh transformation. After fixing the parameters which appear in the wavefunction, Eq. (\ref{FPgeneral}) can be used to fit the experimental data. But this is not the whole story. We consider the term $\widetilde {\phi}_{P} \equiv {\phi_{P}(x,k_\perp)/{\sqrt{{\cal A}^2+k^2_\perp}}}$ and take the Tayor expansion around $k^2_\perp$
\be
\widetilde {\phi}_{P'}(k'^2_\perp)=\widetilde {\phi}_{P'}(k^2_\perp)+{d\widetilde {\phi}_{P'}\over{dk^2_\perp}}\Bigg|_{Q_\perp=0}(k'^2_\perp-k^2_\perp)+{d^2\widetilde {\phi}_{P'}\over{2(dk^2_\perp)^2}}\Bigg|_{Q_\perp=0}(k'^2_\perp-k^2_\perp)^2+.....
\en
Then, by using the idenity
\be
\int d^2k_\perp~(k_\perp \cdot A_\perp)(k_\perp \cdot B_\perp)={1\over{2}}\int d^2k_\perp~k^2_\perp~A_\perp\cdot B_\perp, \label{QQQ2}
\en
we can rewrite (\ref{FPgeneral}) as
\be
F_P(Q^2)&=&(e_{q_1}+e_{\bar q_2})\non \\
&+&Q^2 \int \{dx\}\,\phi^2_P(x,k_\perp)[x^2 e_{q_1}+(1-x)^2e_{\bar q_2}]\Bigg( \Theta_{P}{{\cal A}^2+2k^2_\perp\over{{\cal A}^2+k^2_\perp}}+\widetilde{\Theta}_{P}k^2_\perp\Bigg)\non \\
&+&{\cal O}(Q^4), \label{FPQQ}
\en
where
\be
\Theta_{M}={1\over{\widetilde {\phi}_{M}}}\Bigg({d\widetilde {\phi}_{M}\over{dk^2_\perp}}\Bigg),~~\widetilde{\Theta}_{M}={1\over{\widetilde {\phi}_{M}}}\Bigg({d^2\widetilde {\phi}_{M}\over{(dk^2_\perp)^2}}\Bigg).
\en
From Eq. (\ref{FPQQ}), the static property $F_P(0)=e_P$ is quite easily checked.
The mean square radius of the meson $P$ is determined from the slope of $F_P$ at $Q^2=0$:
\be
\la r^2 \ra_P\equiv -6{dF_P(Q^2)\over{dQ^2}}\Bigg|_{Q^2=0}.
\en 
It should be realized that the size and the density of a hadron depend on the probe. For an electromagnetic probe, it is the electric charge radius that is obtained. From the experimental view, $\la r^2 \ra_P$ cannot be measured directly and is obtained by fitting the data on $F_P$ to a pole or dipole form. Here we easily obtained the equation of $\la r^2 \ra_P$
\be
\la r^2 \ra_P &=& \la r^2 \ra_{q_1}+\la r^2 \ra_{{\bar q}_2} \non \\
&=&e_{q_1}\Bigg\{-6\int\{dx\}x^2\widetilde {\phi}_{P}\Bigg[({\cal A}^2+2k^2_\perp){d\over{dk^2_\perp}}+({\cal A}^2+k^2_\perp)k^2_\perp\Bigg({d\over{dk^2_\perp}}\Bigg)^2\Bigg]\widetilde {\phi}_{P}\Bigg\},\non \\
&+&e_{\bar{q}_2}\Bigg\{-6\int\{dx\}(1-x)^2\widetilde {\phi}_{P}\Bigg[({\cal A}^2+2k^2_\perp){d\over{dk^2_\perp}}+({\cal A}^2+k^2_\perp)k^2_\perp\Bigg({d\over{dk^2_\perp}}\Bigg)^2\Bigg]\widetilde {\phi}_{P}\Bigg\}.\label{MSRP}
\en
From Eq. (\ref{MSRP}), it is worthwhile to mention that, first, the mean square radius  of a meson is the sum of the contributions of the valence quarks. Second, $\la r^2 \ra$ is related to the first and second longitudinal momentum square derivatives of $\widetilde {\phi}$ which contain the Melosh transformation effect.
%%%%%%%%%%%%%%%%%%%%%%%%%section three %%%%%%%%%%%%%%%%%%%%%%%%%%%%%%%%
\section{Heavy Quark Limit}
In this section, we will check the HQS among the charge radii by taking the heavy quark limit.  To proceed, we first investigate the heavy-quark-limit behavior of the wavefunction. Since the $x$ in the normalization condition (\ref{wavenor}) is the longitudinal momentum fraction carried by the light antiquark, the meson wavefunction should be sharply peaked near $x\sim \Lambda_{\rm QCD}/m_Q$. It is thus clear that only terms of the form ``$m_Qx$" survive in the wavefunction as $m_Q\to\infty$; that is, $m_Qx$ is independent of $m_Q$ in the $m_Q\to\infty$ limit. In the $m_Q\to\infty$ limit, we must rewrite Eq. (\ref{wavenor}) in the $m_Q$-independent form
\be
\int^\infty_0dX\int {d^2k_\perp\over 2(2\pi)^3}\left|\Phi(X,k_\perp)
\right|^2=1,  \label{isnom}
\en
where $X\equiv m_Qx$ and \cite{CZL}
\be
\Phi(X,k_\perp)={\phi_{Q_{\bar {q}}}(x,k_\perp)\over{\sqrt{m_Q}}}.  \label{hqamp}
\en
The scaling behavior of Eq. (\ref{hqamp}) is the constraint of the light-front wavefunction when we consider the infinite quark mass limit. For the Gaussian-type wavefunction (\ref{gauss}), it satisfies an asymptotic form
\be
\Phi(X,k_\perp)_{\rm G} =\, 4\left({\pi\over\omega^2}\right)^{3/4}\exp
\left(-{k^2_\perp\over 2\omega^2}\right)\exp\left(-{({X\over 2}-{m_{\bar {q}}^2+
k^2_\perp\over 2X})^2\over 2\omega^2}\right)\sqrt{{1\over 2}+
{m_{\bar {q}}^2+k^2_\perp\over 2X^2}}. \label{GHQ}
\en
Thus we can use this wavefunction when the heavy quark limit is considered.

In the $m_M,m_Q \to \infty$ limit it is appropriate to describe the meson state with the meson velocity $v$ \cite{HQS}
\be
|M(v)\rangle = m_M^{-1/2} |M(P)\rangle,
\en
where $v=P/m_M$. For the decay constant, the definition (\ref{dcd18}) becomes 
\be
\la 0|\bar{q}\gamma_\mu \gamma_5 Q|P(v)\ra &=& \,i\,\bar {f_P}\,v_\mu, \label{dcHV}
\en
and in the $m_Q \to \infty$ limit it is
\be
\bar {f_P}= 4{\sqrt{3}\over{\sqrt{2}}}\int {dX\,d^2k_\perp\over 2(2\pi)^3}\,\Phi(X,k_\perp)\,{\widetilde {\cal A}\over\sqrt{\widetilde {\cal A}^2+k_\perp^2}}, \label{fPV}
\en
where $\widetilde {\cal A} \equiv X+m_{{\bar q}_2}$. Comparing Eq. (\ref{fPV}) with Eq. (\ref{fp}), we obtain the HQS scaling law for the decay constant:
\be
\bar {f_P} = {\sqrt {m_M}} f_P. \label{scale}
\en
For the mean square radius Eqs. (\ref{MSRP}), when the heavy quark limit is considered, we obtain
\be
\la r^2\ra_P = \la r^2\ra_Q+\la r^2\ra_{\bar {q_2}}, \label{rPV}
\en
where
\be
\la r^2 \ra_Q &=&e_Q\Bigg\{{-6\over{m^2_Q}} \int{dXd^2k_\perp\over{2(2\pi)^3}}X^2\widetilde {\Phi}\Bigg[(\widetilde {\cal A}^2+2k^2_\perp){d\over{dk^2_\perp}}+(\widetilde {\cal A}^2+k^2_\perp)k^2_\perp\Bigg({d\over{dk^2_\perp}}\Bigg)^2\Bigg]\widetilde {\Phi}\Bigg\} \non \\
&\to& 0, \label{HQL1} 
\en
\be
\la r^2 \ra_{\bar {q}_2} = e_{\bar {q}_2}\Bigg\{-6 \int{dXd^2k_\perp\over{2(2\pi)^3}}\widetilde {\Phi}\Bigg[(\widetilde {\cal A}^2+2k^2_\perp){d\over{dk^2_\perp}}+(\widetilde {\cal A}^2+k^2_\perp)k^2_\perp\Bigg({d\over{dk^2_\perp}}\Bigg)^2\Bigg]\widetilde {\Phi}\Bigg\},\label{HQL2}
\en
and $\widetilde {\Phi} ={\Phi}/\sqrt{\widetilde {\cal A}^2+k^2_\perp}$. Eq. (\ref{HQL1}) means that the mean square radius $\la r^2 \ra_P$ is blind to the flavor of $Q$. This is the so-called flavor symmetry. We find that the light degrees of freedom are blind to the flavor of the heavy quark. In addition, Ref. \cite{plbhwcw} finds the mean square radius also satisfied the spin symmetry. These are the so-called HQS. Up to now, we have not used the wavefunction yet, this also satisfies the well-known property that HQS is model-independent. Reviewing the processes, we can realize that, in this approach, the static properties of the EM form factors and the heavy quark symmetry of the mean square radii can be checked much more easily than in the Breit frame. This is the major reason why we calculate the $Q^2$ dependence of those form factors order by order.

We must emphasized here that, in the $m_Q \to \infty$ limit, the vanishing of the heavy quark sector in the form factor is true only for the $Q^2 \to 0$ region. In the time-like region, near the threshold for the meson pair production the heavy quark sector is dominant and described by the Isgur-Wise function \cite{MR}.
%%%%%%%%%%%%%%%%%%%%%%%%%section 4 %%%%%%%%%%%%%%%%%%%%%%%%%%%%%%%%
\section{Numerical Results}
In this section, we will use the Gaussian-type wavefunction (\ref{gauss}) to calculate the EM form factors and the mean square radius. The parameters appearing in the wavefunction, the quark mass $m_q$ and the scale parameter $\omega$, are constrained by the decay constants. 

The decay constants of the pseudoscalar mesons $\pi$ and $K$ come from experiments \cite{PDG00}
\be
f_\pi= 130.7~{\rm MeV},~~~ f_K=159.8 ~{\rm MeV},
\en
the others are obtained by lattice and constituent quark model:
\be
f_D &=& 192~{\rm MeV} \cite{lattprl},~~f_{D_s}=210 ~{\rm MeV}\cite{lattprl},~~f_B=157 ~{\rm MeV} \cite{lattprl},\non \\
f_{B_s} &=& 171~{\rm MeV} \cite{lattprl},~~ f_{B_c}=360~{\rm MeV} \cite{QCDSR}.
\en
Combining with the quark masses
\be
m_{u,d}=0.24~{\rm GeV},~~m_s-m_{u,d}=0.18~{\rm GeV},~~m_c=1.6~{\rm GeV},~~m_b=4.8~{\rm GeV}, \label{qmass}
\en
we fit the scale parameters
\be
&&\omega_\pi=0.333~{\rm GeV},~~\omega_K=0.379~{\rm GeV},~~~\omega_D=0.443~{\rm GeV},~~\omega_{D_s}=0.450~{\rm GeV}, \non \\
&&\omega_B=0.477~{\rm GeV},~~\omega_{B_s}=0.485~{\rm GeV},~~\omega_{B_c}=0.813~{\rm GeV}. \label{omega}
\en
There are differences between these parameters and the ones in \cite{QCDSR} because the wavefunctions in the two cases are not the same. However, they have a common tendency such that $\omega_{M_i} < \omega_{M_j}$ if $M_i < M_j$. This corresponds to the ordering law for the size of heavy-light bound states.

The $Q^2$-dependences of $F_\pi$ and $F_K$ can be obtained by Eq. (\ref{FPgeneral}), and we compare the results with the data in Fig. 1 and 2, respectively. In addition, the mean square radii of the pseudoscalar meson can be obtained by Eq. (\ref{MSRP}). We list the results of the $\la r^2 \ra_{\pi^+,K^+,K^0}$ and the experimental data in Table 1. (the unit is $\rm{fm}^2$)

\vskip 0.5cm
\begin{center}
\begin{tabular}{c l l l} \hline
$\la r^2 \ra$ & $\pi^+$ & $K^+$ & $K^0$ \\ \hline
this work & $0.443$ & $0.349$ & $-0.0676$  \\ \hline
[11] & $0.314$ & $0.240$ & $-0.020$ \\ \hline
[13] & $0.452$ & $0.38$ & $0.057$ \\ \hline
experiment & $0.439\pm 0.008$~[1]& $0.34\pm 0.05$~[5] & $-0.054\pm0.026$~[7] \\ \hline
\end{tabular}
\end{center}
\begin{center}
{\small Table 1. The mean square radii of the $\pi^+$, $K^+$, and $K^0$ mesons.}
\end{center}
The negative signs in Table 1 are interesting, and may be interpreted as the preponderance of negative electric charge in the tail of the distribution. We find these values are all consistent with the data. Comparing with Ref. \cite{The0}, they also used the light-front approach. There were various parameter combinations to fit the data of $F_\pi$ for both small and large momentum transfers.

According to the vector meson dominance (VMD) model \cite{The4}, there is a physical explanation: the pion form factor is determined by a $\rho$-meson pole. Generally speaking, this simple picture fits the data well. A detailed study \cite{The5} obtained a better fit when one considers the $\rho$-$\omega$ mixing and three vector meson ($\rho$, $\omega$, and $\phi$) poles to the pion and kaon form factors, respectively.

On the other hand, the mean square radii of the heavy pseudoscalar meson have not been measured yet. For comparsion, here we define and calculate them as $\la r^2 \ra_{FM}$ for the finite quark masses and as $\la r^2 \ra_{IM}$ for the infinite quark masses. In the case of the infinite quark masses, the decay constant $\bar{f_P}$ cannot be measured in the true world, so we obtain it approximately by using the values $f_B=157~{\rm MeV}$ and $m_B=5.28~{\rm GeV}$ in Eq. (\ref{scale}). The results are listed in Table 2. 
\vskip 0.5cm
\begin{center}
\begin{tabular}{c c c c c c c c} \hline
 & $D^+$ & $D^0$ & $D_s^+$ & $B^+$ & $B^0$ & $B_s^0$ & $B_c^+$ \\ \hline
$\la r^2 \ra_{FM}$ & $0.184$ & $-0.304$ & $0.124$ & $0.378$ & $-0.187$ & $-0.119$ & $0.0433$ \\ \hline
$\la r^2 \ra_{IM}$ & $0.248$ & $-0.496$ & $0.181$ & $0.496$ & $-0.248$ & $-0.181$ &  \\ \hline
\end{tabular}
\end{center}
\begin{center}
{\small Table 2. The mean square radii of the heavy pseudoscalar mesons for the finite \\
 quark masses $\la r^2 \ra_{FM}$ and for the infinite quark masses $\la r^2 \ra_{IM}$.}
\end{center}
From Table 2, we cannot obviously find the situation that, comparing with the values in $D_q$ system, the ones in the $B_q$ system are closer to those in the infinite-quark-mass system. The reason is that the $\la r^2 \ra$ is sensitive to the $f_P$, but the uncertainty of the decay constant is not small. In fact, if we use the most recent value $f_{D_s}=280~{\rm MeV}$ \cite{chada}, the result $\la r^2 \ra_{D^+_s}=0.083~{\rm fm}^2$ is quite different from the one in Table 2. For the $B_c$ meson, the $\la r^2 \ra_{IM}$ are not given here because both $b$ and $c$ quarks are heavy. The HQS must be reconsidered in this case.
%%%%%%%%%%%%%%%%%%%%%%%%%section 5 %%%%%%%%%%%%%%%%%%%%%%%%%%%%%%%%
\section{Conclusion}
We have calculated the EM form factors of the pseudoscalar mesons. The EM form factors are extracted from the relevant matrix elements directly, instead of choosing the Breit frame. We found that the charge radius is related to both the first and second longitudinal momentum square derivatives of the momentum distribution function. We also found that the static properties of the EM form factors and the heavy flavor symmetry of the mean square radii are checked analytically by evaluating the $Q^2$ dependence of those form factors order by order. Therefore, in the heavy quark limit, the charge radii of pseudoscalar have flavor symmetries, and these properties are model-independent. In addition, The $Q^2$-dependences of the form factors $F_{\pi,K}$ and the mean square radius of light and heavy mesons have been calculated by using the Gaussian-type wavefunction. The form factors $F_\pi$ and $F_K$ in small momentum transfer and the values of $\la r^2 \ra_{\pi^+,K^+,K^0}$ are all consistent with the current experimental data.
\vskip 1.0cm
\centerline {\bf ACKNOWLEDGMENTS}

I would like to thank Kingman Cheung for helpful comments. This work was supported in part by the National Science Council of ROC under Contract No. NSC90-2112-M-007-040.

%%%%% References %%%%%%%%%%%%%%%
%
%
\newcommand{\bi}{\bibitem}
%

%%%%% Figure Captions %%%%%%%%%%%
\newpage
\parindent=0 cm
\centerline{\bf FIGURE CAPTIONS}
\vskip 0.5 true cm

{\bf Fig. 1 } The charge form factor of the pion in small momentum transfer. Data are taken from \cite{Exp0}.
\vskip 0.25 true cm

{\bf Fig. 1 } The charge form factor of the Kaon in small momentum transfer. Data are taken from \cite{Exp4}.
\vskip 0.25 true cm
\newpage

\begin{figure}[h]
\hskip 4cm
\hbox{\epsfxsize=15cm
      \epsfysize=12cm
      \epsffile{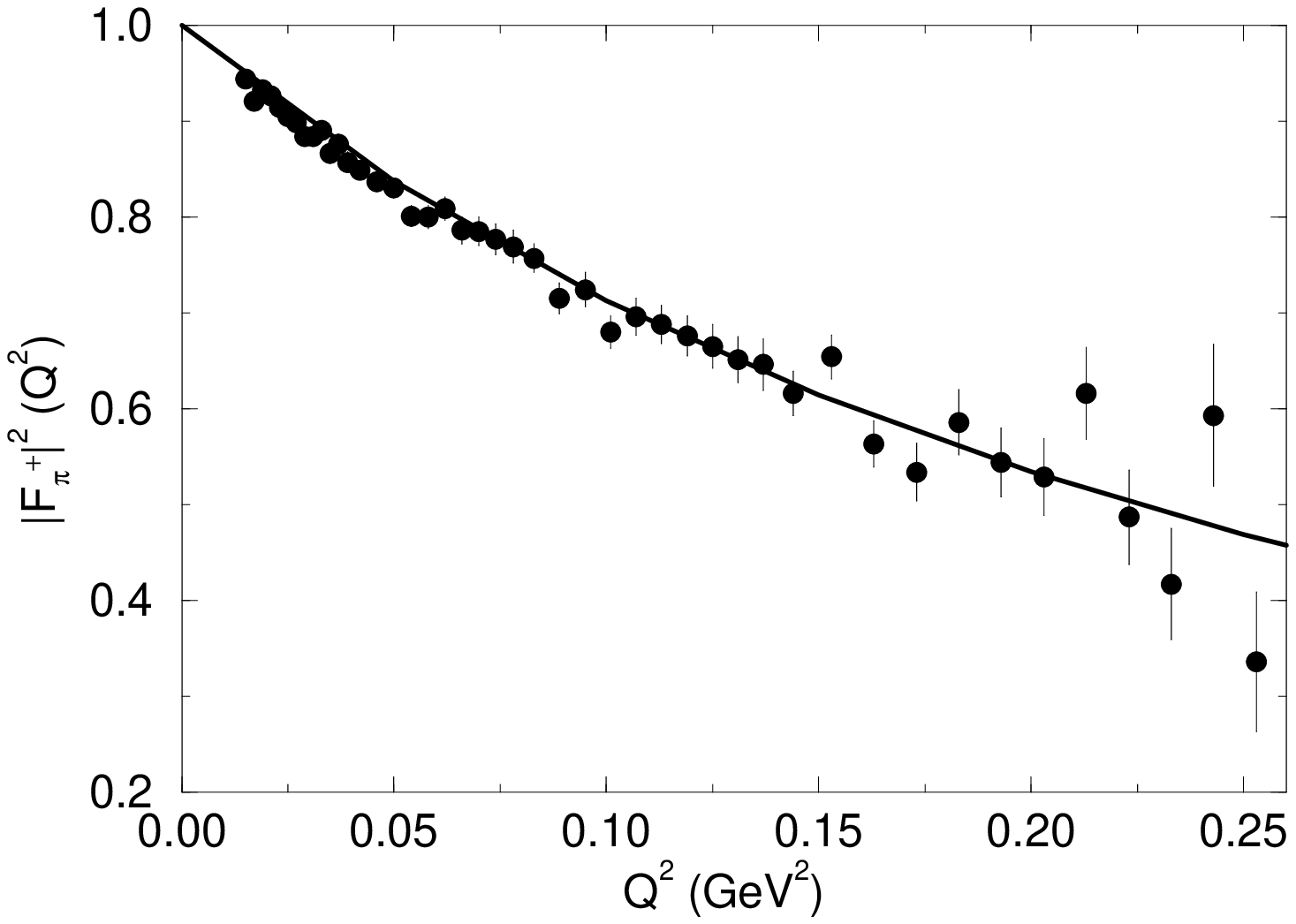}}
\end{figure}
\newpage

\begin{figure}[h]
\hskip 4cm
\hbox{\epsfxsize=15cm
      \epsfysize=12cm
      \epsffile{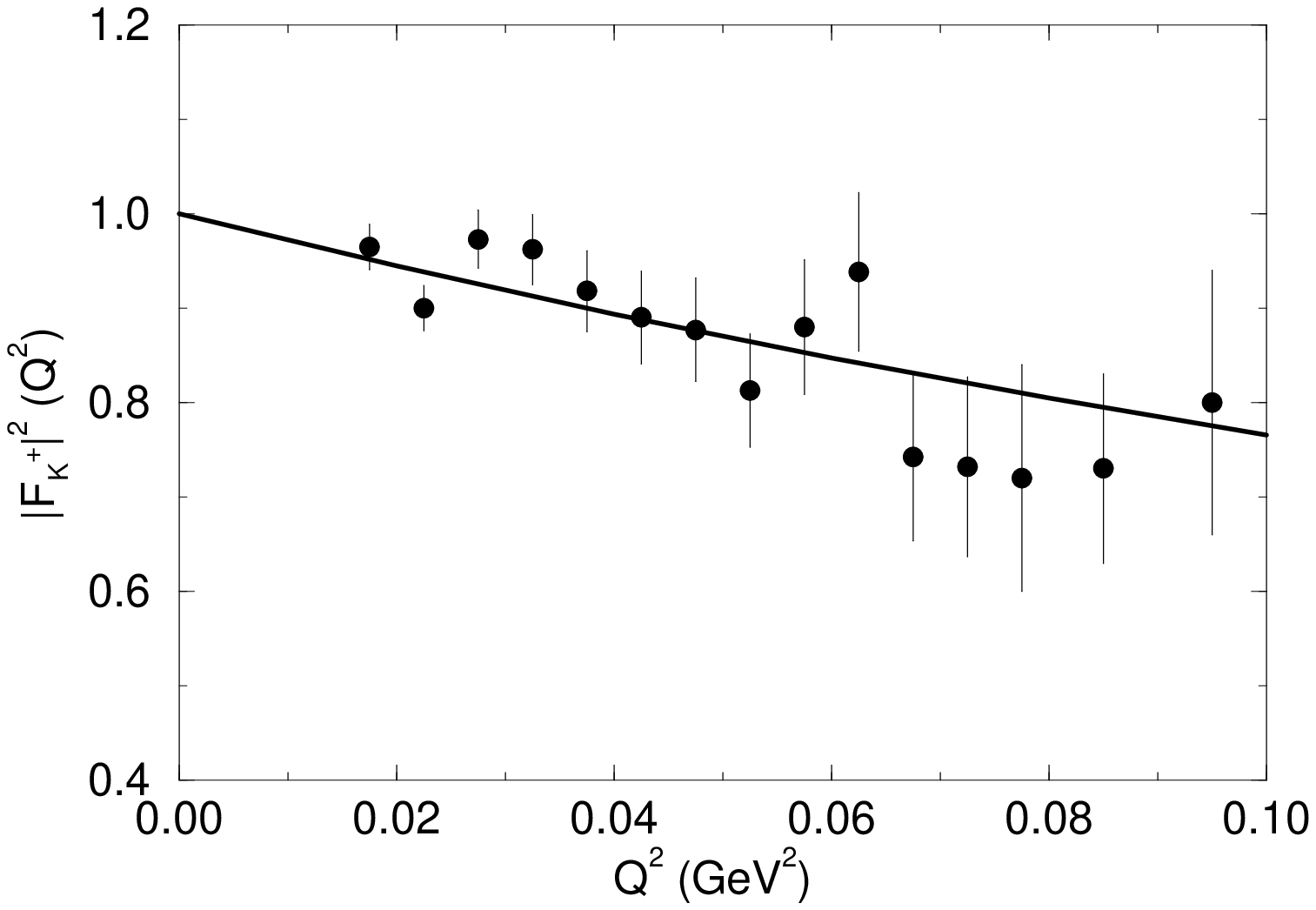}}
\end{figure}
\newpage
\end{document}